\documentclass[a4paper,10pt]{notices}
\usepackage{amsfonts,amssymb,amsmath,amscd}
\usepackage{natbib}
\usepackage[most]{tcolorbox}
\usepackage[greek,english]{babel}

\usepackage{bm}
\usepackage{microtype}
\usepackage[compact,small]{titlesec}
\usepackage{type1cm}
\newlength\normalFontSize\newlength\fontSizeIncrement
\usepackage{xcolor}
\definecolor{titlegray}{RGB}{105,105,105} 

\definecolor{amsblue}{HTML}{1E5AA6}

\title{\Huge\color{amsblue}{A Brief History of Inference in Astronomy}}

\author{
  \Large \it Rafael S. de Souza
  \affil{
    Rafael S. de Souza is a Senior Lecturer in Data Science at the University of Hertfordshire. His email address is rd23aag@herts.ac.uk.
    }
  \and
   \Large \it Emille E. O. Ishida
  \affil{
    Emille E. O. Ishida is a CNRS Research Engineer at Universit\'e Clermont Auvergne. Her email address is emille.ishida@clermont.in2p3.fr
   }
 \and
  \Large \it Alberto Krone-Martins
  \affil{
    Alberto Krone-Martins is a Lecturer in the Department of Informatics at the University of California Irvine. His email address is algol@uci.edu. 
   }
}

\begin{document}

\maketitle

In this short review, we trace the evolution of inference in astronomy, highlighting key milestones rather than providing an exhaustive survey. 
We focus on the shift from classical optimization to Bayesian inference, the rise of gradient-based methods fueled by advances in deep learning, and the emergence of adaptive models that shape the very design of scientific datasets. Understanding this shift is essential for appreciating the current landscape of astronomical research and the future it is helping to build (see also \citet{Feigelson2012,Tak_2024}).

Inference—the process of drawing conclusions about a model or its parameters from data—is central to scientific inquiry into the Universe. Since ancient times, humans have built what we now call  \emph{predictive models}, aimed at estimating future outcomes under given conditions in the world, and \emph{explanatory models}, to elucidate relationships among variables and the mechanisms underlying observed phenomena. In all cases, models depend on parameters to be inferred or distributions to be determined. Throughout history, methods of inference have been deeply linked to advances in astronomy, dating back to Babylonian times  \citep{Plackett1958,Stigler1990}.

According to \selectlanguage{greek}Πτολεμαῖος\selectlanguage{english} (Ptolemy), in his 2\textsuperscript{nd}-century AD treatise \selectlanguage{greek}{Μαθηματικὴ Σύνταξις} \selectlanguage{english}(later known through Arabic translations as the Almagest), the Greek astronomer \selectlanguage{greek}Ἵππαρχος\selectlanguage{english} (Hipparchus) improved the precision of his astronomical measurements—such as the timing of solstices, equinoxes, and lunar eclipses—by averaging the earliest and latest observations of the same event. This approach mitigates observational noise by assuming that timing errors are roughly symmetric, allowing deviations to cancel and yielding a more accurate estimate of the true event time.

In the 16\textsuperscript{th} century, the Portuguese mathematician and cosmographer Pedro Nunes discovered the loxodrome, or rhumb line,\footnote{In navigation, a rhumb line is a path crossing all meridians at the same angle. See: \url{https://en.wikipedia.org/wiki/Rhumb_line}}
({\it linha de rumo}, in Portuguese),
on the surface of a sphere and, perhaps more importantly, invented the Nonius\footnote{\url{https://en.wikipedia.org/wiki/Nonius_(device)}}. 
By combining readings from several closely spaced, independent scales, the Nonius enabled more precise measurements of quantities such as the altitude angles of celestial bodies \citep{PedroNunes1542}.

A century later, a simplified version of Nunes' method gave rise to the Vernier scales still used in modern calipers. With their fine divisions, these scales remain one of the key innovations behind the precision of modern absolute encoders, integral to computer numerically controlled machines and other robots essential to contemporary life. Nunes' scales and methods inspired Tycho Brahe, as Tycho himself acknowledged in his book \textit{Astronomiæ instauratæ mechanica} \citep[Instruments for the restoration of astronomy;][]{1602tbam.book.....B}. 
In the 16\textsuperscript{th} century, Tycho Brahe appears to have been the first to deliberately make a large number of repeated observations of the same quantity—specifically, the positions of stars—which he later combined into a single inferred value \citep{dreyer2004tycho}.
This process enabled Tycho to achieve unprecedented precision in determining the coordinates of individual stars and the motions of planets, laying the groundwork for Kepler's laws of planetary motion—an achievement that definitively displaced Earth from the center of the Universe.

Copernicus’s 1543 heliocentric model, \textit{De revolutionibus orbium coelestium},  had already challenged Ptolemy’s system, but it was Kepler’s laws, backed by Tycho’s precise data, that sealed the case. Not only did they rest on better observations, but they embodied what William of Ockham had axiomatized in his \textit{Summa Logicae}: the principle of parsimony, or ``Occam’s razor".  In practice, this meant that fewer adjustable parameters were needed to match observation, lending the heliocentric picture a decisive methodological advantage over its geocentric rival.

Kepler’s laws, inferred directly from Tycho’s precise data, served as predictive models. Their physical explanation later inspired Newton’s laws of motion and gravity, along with the development of infinitesimal calculus \citep{1687pnpm.book.....N, POURCIAU200118}. In the same century, Galileo's 1609 \textit{Dialogo sopra I due massimi sistemi del mondo} (Dialogue on the Two Great World Systems) offered an early discussion of observational errors concerning the distance estimate to the 1572 supernova in the constellation of Cassiopeia. The formal understanding of how multiple observations improve inference would emerge only later.

In his \textit{Théorie analytique des probabilités} (1812), Pierre-Simon Marquis de Laplace extended  Abraham De Moivre’s formulation of the result later named the \textit{zentraler Grenzwertsatz} (Central Limit Theorem) by George Pólya. De Moivre first introduced it in a  1733 article and later included it in the \textit{The Doctrine of Chances} (1738). This cornerstone of probability theory was further refined by Augustin-Louis Cauchy, Siméon-Denis Poisson, and Friedrich Wilhelm Bessel \citep{Stahl06}.

In the same century, Carl Friedrich Gauss gained early recognition in astronomy by accurately predicting the orbit of the planetoid Ceres, briefly observed in early 1801 before disappearing. Ceres was located near Gauss’s predicted position later that year. He published his methods in {\it Theoria motus corporum coelestium}, \citep{gauss1809}, introducing the method of least squares,  which he claimed to have developed in 1795. This led to a controversy with Adrien-Marie Legendre, who had published a similar least squares method in his 1805 work \textit{Nouvelles méthodes pour la détermination des orbites des comètes}. Regardless of this dispute, Laplace was the first to formalize the method as a probabilistic problem in his 1812 {\it Théorie Analytique des Probabilités}.

Curiously, inverse probability and Bayes' theorem, developed by Thomas Bayes and Pierre-Simon Laplace from Laplace’s own astronomical work, had limited influence in astronomy until the late  20\textsuperscript{th} century.  The broader adoption of Bayesian methods in the field owes much to Sir Harold Jeffreys, a geophysicist and Plumian Professor of Astronomy at Cambridge, who axiomatized Laplace’s theory in his 1939 book, Theory of Probability.
By the 1990s, Bayesian methods saw broader use, especially in extragalactic astronomy and cosmology \citep[e.g.][]{Loredo1992,Molina1992}. The rise of advanced Bayesian techniques and probabilistic programming is now enabling astronomers to address increasingly complex problems.


\section*{Linear Regression in Astronomy}

Linear regression has long been central to astronomical data analysis \citep{Feigelson2012}, dating back to its early use by Isaac Newton around 1700 to study the timing of the equinoxes -- points in Earth's orbit when the Sun crosses the celestial equator. He achieved this by combining ancient data from Hipparchus with observations by his contemporary, John Flamsteed \citep{2016Obs...136....1B}.

An archetypal example includes early attempts to estimate the best-fit line in Hubble’s diagram, which relates galaxies’ recessional velocities (how fast they are moving away from us) to their distances. These analyses revealed that the Universe is expanding, with the slope of the line defining the Hubble constant $H_0$, a measure of the current rate of cosmic expansion \citep{1927ASSB...47...49L,hub29}. Formally, \( H_0 = \left.\dfrac{\dot{a}(t)}{a(t)}\right|_{t = t_0} \), where \( a(t) \) is the scale factor describing how distances in the Universe evolve with time, and \( t_0 \) is the present cosmic time. Intuitively, if \( H_0 \approx 70\, \mathrm{km\,s^{-1}\,Mpc^{-1}} \), remained constant, the size of the Universe would double approximately every 10 billion years.\footnote{1 parsec (pc) = \(3.086 \times 10^{13}\) km; 1 Mpc = \(10^6\) pc.}


Among its many applications, linear regression also played a pivotal role in Hulse and Taylor’s Nobel Prize–winning research on binary pulsars \citep{Taylor1982}. The Hulse–Taylor pulsar (PSR~B1913+16) was the first binary pulsar  discovered,  a system of two neutron stars orbiting their common center of mass. Neutron stars are  remnants of massive stars, typically about a dozen kilometers in diameter, whose cores collapsed to densities exceeding that of an atomic nucleus. In such a system, one of the stars appears as a highly regular radio pulsar: a rapidly spinning neutron star that emits beams of radio waves from its magnetic poles. These beams sweep across Earth at regular intervals, much like a lighthouse, allowing precise timing measurements. The companion star is often electromagnetically invisible  and detectable only through its gravitational effects.  Observations of orbital decay in the system closely matched  Einstein’s prediction of energy loss via gravitational wave emission, providing indirect evidence for gravitational waves more than two decades before their direct detection.\footnote{The ratio of the observed to predicted orbital decay is 0.997 $\pm$ 0.002 \citep{Weisberg2010}.}

Beyond compact-object systems, linear regression also plays a crucial role in uncovering large-scale empirical relations in extragalactic astrophysics. Many such scaling laws, often expressed as power laws, can be linearized by taking logarithms. A classical example is the log-linear relationship between the mass of a galaxy's central supermassive black hole (\(M_{\bullet}\)) and the stellar velocity dispersion (\(\sigma_{e}\)) in its bulge. The bulge is the dense, roughly spherical central region, composed primarily of older stars. The stellar velocity dispersion measures the spread in stellar velocities within the bulge. This relationship suggests a feedback mechanism: the growth of the black hole and the evolution of its host galaxy are interconnected—possibly through energetic outflows from the black hole that regulate star formation in the surrounding galaxy.

This empirical relation is expressed as:
\begin{equation}
\log M_{\bullet} = \alpha + \beta \log \sigma_{e} + \epsilon, \quad \epsilon \sim \mathcal{N}(0, \sigma^2),
\end{equation}
where \(\alpha\) and \(\beta\) are the regression coefficients, and \(\sigma^2\) is the intrinsic scatter variance. Parameter estimation often reduces to an inverse problem, with least-squares serving as a foundational tool as it has ever since Gauss’s determination of the orbit of Ceres \citep{gauss_ceres}.

These early applications exemplify a broader principle: many problems in astronomy reduce to estimating parameters that best explain observational data. Whether fitting a line to a Hubble diagram or deriving black hole scaling relations, the core task often involves comparing model predictions with observations. A common formulation of this task is the estimation of model parameters \(\theta\) by comparing theoretical predictions \(\mu(\theta)\) to observed data \(\mathbf{d}\).  This naturally leads to a weighted least squares problem. 

Assuming independent, identically distributed (i.i.d.) Gaussian errors, one seeks to minimize the discrepancy between model and data:

\begin{equation}
\mathcal{S}(\theta) = (\mathbf{d} - \mu(\theta))^\top \mathbf{W} (\mathbf{d} - \mu(\theta)),
\end{equation}
where \(\mathbf{W}\) is a diagonal matrix of inverse variances representing the measurement uncertainties. The best-fit parameters \(\hat{\theta}\) are those that minimize this objective function:
\begin{equation}
\hat{\theta} = \arg \min_{\theta} \mathcal{S}(\theta).
\end{equation}
Popular optimization techniques include the downhill simplex method \citep{Nelder65}, 
Levenberg--Marquardt \citep{marquardt1963algorithm},
conjugate gradient \citep{Hestenes1952MethodsOC}, 
and gradient descent.  The latter updates parameter estimates iteratively according to:
\begin{equation}
\theta^{(r+1)} = \theta^{(r)} - \alpha \nabla \mathcal{S}(\theta^{(r)}), 
\end{equation}
where \(\alpha\) is the step size, and $\nabla \mathcal{S}(\theta^{(r)})$ is the gradient of the objective function evaluated at the current parameter estimate.

When latent variables or incomplete data are present, the 
\emph{expectation--maximization} algorithm  \citep{Dempster1977,Wu1983} is often preferred for maximum likelihood estimation.   The likelihood function \(\mathcal{L}(\theta)\) represents the probability mass (for discrete outcomes) or density (for continuous ones) of the observed data, regarded as a function of the parameters \(\theta\), with the data held fixed. The goal is to find
\begin{equation}
\hat{\theta} = \arg \max_{\theta} \mathcal{L}(\theta),
\end{equation}
which can also be approached using gradient-based updates:
\begin{equation}
\theta^{(r+1)} = \theta^{(r)} + \alpha \nabla \mathcal{L}(\theta^{(r)}).
\end{equation}
Both EM and gradient ascent share a common iterative structure:
\begin{equation}
\psi^{(r+1)} = \mathcal{U}(\psi^{(r)}),
\end{equation}
where \(\psi\) denotes the model parameters, and \(\mathcal{U}\) is an update operator that increases the likelihood or decreases an objective function at each step.

\begin{figure}
    \centering
    \includegraphics[width=1\columnwidth]{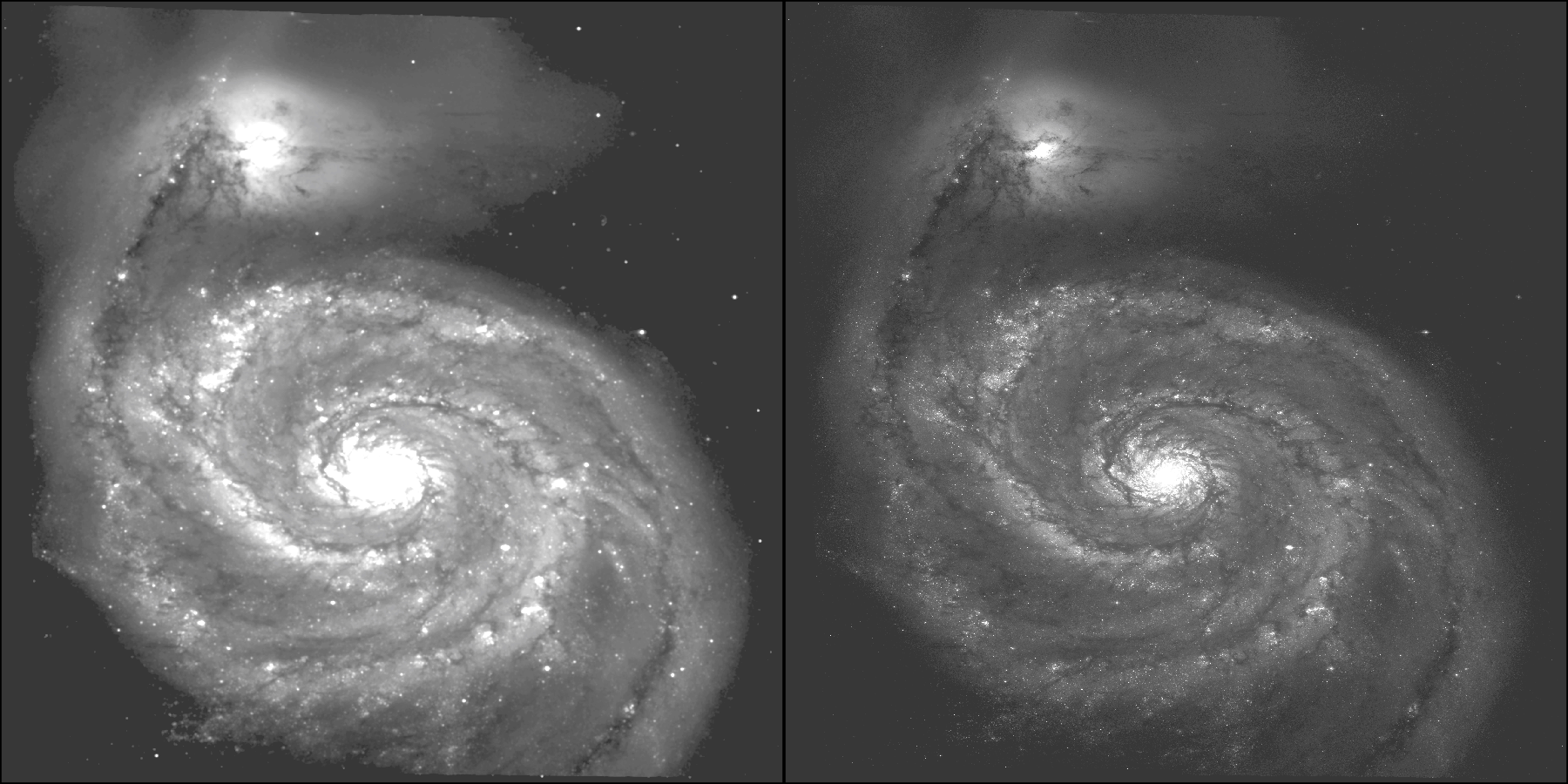}
    \caption{Left: A blurred, Poisson-noised image of the Whirlpool Galaxy M51. Right: Denoised version obtained via the Richardson–Lucy deconvolution. The method enhances the contrast of fine details such as spiral arms. Credit: NASA, ESA/Hubble, and the Hubble Heritage Team (STScI/AURA).}
    \label{fig:RL}
\end{figure}

A classical example in astronomical image reconstruction is the
\emph{Richardson--Lucy deconvolution} \citep{Richardson1972, Lucy1974}, which is used to  recover images degraded  by the point-spread function (PSF)\footnote{The PSF describes how a point source is blurred by the telescope, detector, or atmosphere.} under Poisson noise.
The observed image \(H_k\) is modeled as a Poisson realization whose mean is the 
convolution of the true image \(W_j\) with the PSF \(P_{kj}\):
\begin{equation}
H_k \sim \mathrm{Poisson}\Big( \sum_j P_{kj} W_j \Big). 
\end{equation}
The corresponding likelihood function is:
\begin{equation}
\mathcal{L}(W) = \prod_k \frac{\left( \sum_j P_{kj} W_j \right)^{H_k} 
e^{ - \sum_j P_{kj} W_j }}{H_k!}.
\end{equation}
Applying the expectation--maximization algorithm to this model leads to the classical Richardson--Lucy update rule:
\begin{equation}
W_j^{(r+1)} = W_j^{(r)} \sum_k \frac{H_k}{\sum_{j'} P_{kj'} W_{j'}^{(r)}} P_{kj}.
\end{equation}
Figure~\ref{fig:RL} shows the Richardson–Lucy deconvolution  applied to a blurred, Poisson-noised image of the Whirlpool Galaxy (M51). The left panel shows the degraded image, while the right panel displays the deconvolved result, with finer structures—such as spiral arms—becoming more discernible. Note that although motivated by Bayesian reasoning, the Richardson–Lucy algorithm produces a maximum likelihood estimation rather than a full posterior distribution. As with other EM and gradient-based approaches, it iteratively refines parameter estimates by maximizing the (log-)likelihood or, equivalently, minimizing an objective function.

Starting with the next section, we follow a focused cosmological example as an illustrative case to trace how this conceptual framework has evolved in response to increasingly complex data and models.

\section*{Measuring the Universe Expansion Rate}

As noted earlier, in the early decades of the 20\textsuperscript{th} century, Edwin Hubble provided the first observational evidence for cosmic expansion by showing that the recession velocities of extra-galactic nebulae are proportional to their distance. These nebulae were later recognized as galaxies beyond the Milky Way.  Hubble inferred their velocities from their redshifts, the Doppler-induced shift of spectral lines toward longer wavelengths, and thus established that more distant objects move away faster \citep{hub29}.

\begin{figure}
    \centering
    \includegraphics[width=\columnwidth]{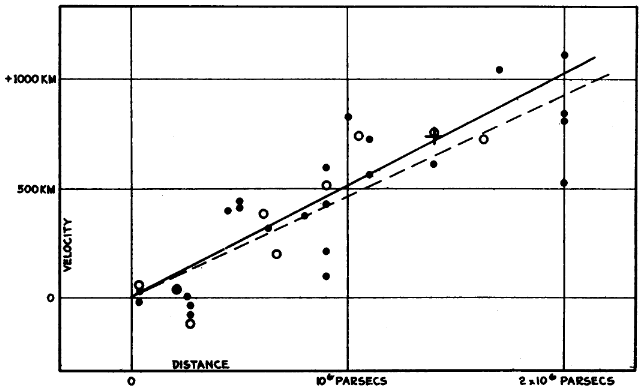}
    \caption{Original plot of radial velocity-distance relation for extragalactic nebulae \citep{hub29}.}
    \label{fig:hubble1929}
\end{figure}

This conclusion emerged from a linear regression analysis of measurements from 24 nebulae in our cosmic neighborhood (Figure~\ref{fig:hubble1929}). It had a profound impact on subsequent cosmological studies, sparking a century-long effort to collect increasingly precise data in the hope of getting more clues about the overall dynamics and composition of cosmic structures.

Nearly a century later, Type Ia supernovae provided the first compelling evidence for the Universe's accelerated expansion \citep{riess1998, perlmutter1999}. This led to the discovery of dark energy—a mysterious form of energy that permeates space and drives cosmic acceleration. The 2011 Nobel Prize in Physics was awarded for this discovery.

Type Ia supernovae mark the catastrophic endpoint in the life of a binary star system, with a white dwarf at the core of the process. In such systems, two stars orbit a shared center of mass, and over time, the white dwarf, a compact remnant left behind by a Sun-like star, can accrete material from its companion. White dwarfs are supported against gravitational collapse by electron degeneracy pressure, a quantum mechanical effect arising from the Pauli exclusion principle, which prevents electrons from occupying the same quantum state. As the white dwarf accumulates mass, it eventually approaches the Chandrasekhar limit (about 1.4 times the mass of our Sun), beyond which degeneracy pressure can no longer support it. This triggers  a runaway thermonuclear reaction that tears the star apart in a brilliant supernova explosion.

These explosions are among the most energetic events in the Universe, releasing an enormous yet remarkably consistent amount of energy at peak brightness (or \emph{luminosity}, the total energy emitted per unit time). This near-uniformity enables Type Ia supernovae to serve as \emph{standardizable candles}—astronomical objects whose intrinsic brightness can be empirically calibrated. By comparing the observed brightness to this calibrated intrinsic brightness, astronomers can infer distances, making Type Ia supernovae invaluable objects  for measuring cosmic distances on large scales \citep{liu2023}.

This standardization is achieved through empirical corrections to the observed magnitude that account for variations in the light curve—the evolution of a supernova’s brightness over time—particularly its width (known as stretch) and its color (which reflects shifts in the observed spectrum toward shorter or longer wavelengths). These corrections substantially reduce the intrinsic scatter in the magnitude–distance relation and improve the precision of cosmological distance estimates \citep{phillips1993}.

The observed magnitude \(m_{\rm obs}\) of a Type Ia supernova at maximum brightness, a logarithmic measure of its apparent brightness as seen from Earth is related to the distance modulus \(\mu\) by:
\begin{equation}
m_{\rm obs} = \mu + M_{\rm int}  \,\,\underbrace{-\alpha s + \beta c}_{\mathcal{O}(s, c)},
\end{equation}
where \(M_{\rm int}\) is the absolute magnitude, defined as the magnitude the object would have if placed at a reference distance of 10 parsecs. Both observed and absolute magnitudes are logarithmic measures of brightness, with lower values corresponding to brighter objects. The parameters \(s\) and \(c\) represent the light-curve stretch and color, respectively, while \(\alpha\) and \(\beta\) are empirical coefficients.  For illustrative purposes, we will henceforth ignore the corrections $\mathcal{O}(s, c)$, as well as measurement uncertainties.

The distance modulus \(\mu\) quantifies the relationship between intrinsic and observed brightness as a function of distance:
\begin{equation}
\mu = 5 \log_{10}\left(\frac{d_L(z)}{10 \, \text{pc}}\right),
\end{equation}
where \(d_L\) is the luminosity distance, given by:
\begin{equation}
d_L(z) = (1+z)\frac{c}{H_0} \int_0^z \frac{dz'}{E(z')}.
\label{eq:dl}
\end{equation}
Here, \(z\) is the redshift, \(c\) is the speed of light and $H_0$ is the Hubble constant, which denotes the relative rate of current cosmic  expansion.  While \(d_L\) is expressed in parsecs in equation \ref{eq:dl}, it is common in observational cosmology to use  distances in Mpc, in which case \(\mu\) becomes:
\begin{equation}
\mu = 5 \log_{10}\left(d_L(z)\right) + 25. 
\end{equation}

Assuming a flat Universe containing only dark matter and dark energy, the luminosity distance $d_L$ for a source with redshift $z$, depends on the dark matter density ($\Omega_m$) and the dark energy equation of state ($w$) parameters through 
\begin{equation}
E(z) = \sqrt{\Omega_m(1 + z)^3 + (1 - \Omega_m)(1+z)^{3(1+w)}}.
\end{equation}
Here, dark matter refers to a non-luminous form of matter that interacts gravitationally but neither emits nor absorbs light. It plays a crucial role in the formation of cosmic structures. The assumption of flatness corresponds to a Universe with zero spatial curvature, as predicted by inflationary cosmology and supported by measurements of the cosmic microwave background. In general, a positively curved (closed) Universe would cause light rays to converge, while a negatively curved (open) Universe would cause them to diverge, altering the relationship between $z$ and $d_L(z)$.

The likelihood for the observed magnitudes  \(m_{\rm obs}\) is given by:
\begin{equation}
\mathcal{L}(\theta | m_{\rm obs}) = \prod_{i=1}^N \frac{1}{\sqrt{2\pi \epsilon^2}} \exp\left( -\frac{(m_{\textrm{obs},i} - \eta_i)^2}{2\epsilon^2} \right),
\end{equation}
where
\begin{equation}
\eta_i = 25 + 5\log_{10}\left(d_L(H_0, w, \Omega_m)\right) + M_{\rm int}.
\end{equation}
Here, \(\epsilon\) is the uncertainty in the observed magnitudes, \(N\) is the total number of observed Type Ia supernovae, and the parameter vector is \(\theta = (H_0, w, \Omega_m, M_{\rm int})\). The corresponding log-likelihood is:
\begin{equation}
\ln \mathcal{L}(\theta | m_{\rm obs}) = -\frac{N}{2} \ln(2\pi \epsilon^2) - \frac{1}{2\epsilon^2} \sum_{i=1}^N (m_{\textrm{obs},i} - \eta_i)^2.
\end{equation}

Estimating cosmological parameters, such as \(w\) and \(\Omega_m\), along with their potential variations, has been a driving force behind the development of a range of observational strategies and large-scale astronomical surveys. 
These include weak gravitational lensing, the subtle distortion of galaxy shapes due to the bending of light by intervening mass; baryon acoustic oscillations (periodic fluctuations in galaxy clustering that serve as a cosmic distance scale), gravitational waves (ripples in spacetime produced by cataclysmic events like merging black holes),  galaxy clusters (whose abundance constrains structure formation),  and the cosmic microwave background, the relic radiation from the early Universe that encodes information about its initial conditions and composition. 

To harness the full potential of these ever-growing datasets, astronomers have increasingly adopted advanced statistical methods—ranging from sampling techniques such as Hamiltonian Monte Carlo and nested sampling to modern approaches in likelihood-free and amortized inference. The following sections examine how these tools are reshaping cosmological analysis, particularly when likelihoods are intractable, simulations costly, or adaptive observational strategies required.





\subsection*{Hamiltonian Monte-Carlo}

Markov chain Monte Carlo (MCMC) has been a cornerstone of computational Bayesian inference for decades. Two widely used methods in astronomy are the Metropolis-Hastings \citep{metropolis1949monte} algorithm and Gibbs sampling \citep{Geman1984}. In the case of Metropolis-Hastings, each MCMC step proposes a new state \(\theta'\) from a proposal distribution \(q(\theta'|\theta_t)\), which is then accepted  with a probability determined by the posterior ratio. The use of MCMC methods in cosmology has grown since the early 21\textsuperscript{st} century, supported by increasing availability of computational power. 
These methods are particularly useful for 
cosmological parameter estimation.

Hamiltonian Monte Carlo \citep[HMC;][]{DUANE1987216} improves sampling efficiency over Metropolis-Hastings by reducing autocorrelation and avoiding the inefficiencies of random walk behavior. It is particularly effective for high-dimensional and strongly correlated posterior distributions, which are common in astrophysical applications.

The main idea behind HMC is to reinterpret the negative log-posterior as a \emph{potential energy} function.  The parameter space \(\theta \in \mathbb{R}^d\) is augmented with auxiliary \emph{momentum variables} \(\phi \in \mathbb{R}^d\), and their joint dynamics are governed  by the \emph{Hamiltonian}, which represents the total energy of the system:
\begin{equation}
H(\theta, \phi) = U(\theta) + K(\phi),
\end{equation}
where the potential energy is
\begin{equation}
U(\theta) = -\log p(\theta | D),
\end{equation}
and the kinetic energy is
\begin{equation}
K(\phi) = \frac{1}{2} \phi^\top M^{-1} \phi,
\end{equation}
with \(M\)  a mass matrix, often set to the identity.

\begin{figure}
    \centering
    \includegraphics[width=\columnwidth]{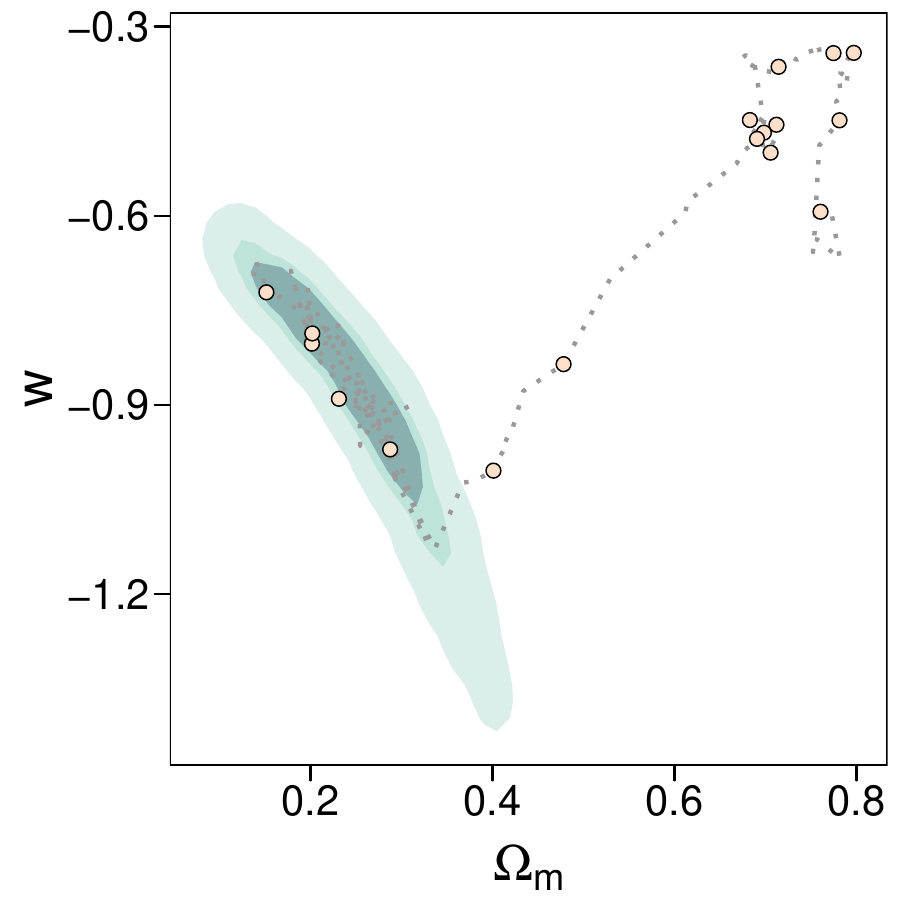}
    \caption{Hamiltonian Monte-Carlo trajectory.}
    \label{fig:fit}
\end{figure}

This physical analogy enables the sampler to explore the posterior by simulating the motion of a particle through parameter space, following smooth, energy-conserving trajectories. In contrast to the random-walk proposals used in traditional Markov chain Monte Carlo, Hamiltonian Monte Carlo generates distant, low-autocorrelation proposals that more efficiently traverse the high-probability regions of complex, high-dimensional posteriors \citep{Betancourt2017}. Figure~\ref{fig:fit} illustrates this intuition in a cosmological context, showing a typical Hamiltonian Monte Carlo trajectory when estimating \(w\) and \(\Omega_m\).

\section*{Amortized Inference}

Astronomical research often relies on computationally intensive simulation to model physical processes that lie beyond the reach of direct observation. These include hydrodynamical, radiative, and magnetohydrodynamical simulations that aim to reproduce the complex behavior of astrophysical systems. However, performing full Bayesian inference using Markov chain Monte Carlo methods is often computationally prohibitive, especially with datasets comprising millions or billions of observations.

To address these challenges, astronomers have increasingly adopted scalable inference techniques, such as Approximate Bayesian Computation (ABC). Initially developed in population genetics and rooted in Rubin's \citep{Rubin1984} rejection sampling thought experiment, ABC has since been adapted for astronomical applications \citep{tavare1997inferring,pritchard1999population}. Over the past decade, ABC has gained traction for tackling problems like measuring the Universe’s expansion rate \citep{Weyant2013}, estimating cosmological parameters from galaxy clusters number counts \citep{ISHIDA20151} and modeling galaxy populations \citep{Cameron2012}.

Recent efforts have increasingly focused on \emph{amortized inference}, a class of  \emph{simulation-based} or \emph{likelihood-free} methods that leverage learned surrogate models -- typically neural networks -- to approximate posterior distributions.  Unlike approximate Bayesian computation, which performs inference separately for each dataset, amortized approaches learn a global mapping from observations to posteriors that can be reused across datasets.

Formally, let \(p(\theta, D)\) define a probabilistic model with latent parameters $\theta \in \mathcal{X}$ and observations $D \in \mathcal{Y}$, where 
$\mathcal{X}$ and $\mathcal{Y}$ denote the latent and data spaces, respectively. The goal is to learn a mapping $g_\phi: \mathcal{Y} \to \mathcal{P}(\mathcal{X})$, where \(g_\phi(D)\) defines an approximate posterior distribution \(q_\phi(\theta | D)\), parameterized by \(\phi\). Learning proceeds by minimizing the discrepancy between the true posterior \(p(\theta|D)\) and the approximate posterior \(q_\phi(\theta | D)\), typically via the Kullback-Leibler ($\mathrm{D_{KL}}$) divergence:

\begin{equation}
\phi^* = \arg\min_\phi \mathbb{E}_{D \sim p(D)} \Bigl[ \mathrm{D_{KL}} \left( p(\theta | D)  \Bigm\| q_\phi(\theta | D) \right) \Bigr].
\end{equation}

In simulation-based inference, the true posterior \(p(\theta | D)\) is typically intractable, making the Kullback-Leibler divergence objective a conceptual ideal rather than directly computable. In practice, this is addressed via methods such as \emph{neural posterior estimation} (which learns \(q_\phi(\theta | D)\) directly), \emph{neural likelihood estimation} (which models \(p(D | \theta)\)),  \emph{neural ratio estimation} (which estimates ratios like \(r(D, \theta) = \frac{p(D | \theta)}{p(D)}\)), and adaptations of \emph{variational inference} for likelihood-free settings. Although these approaches rely on simulations to train surrogate models, they do not generally optimize the $\mathrm{D_{KL}}$ divergence explicitly \citep{Cranmer2020}.

\begin{figure}
    \centering
    \includegraphics[width=\columnwidth]{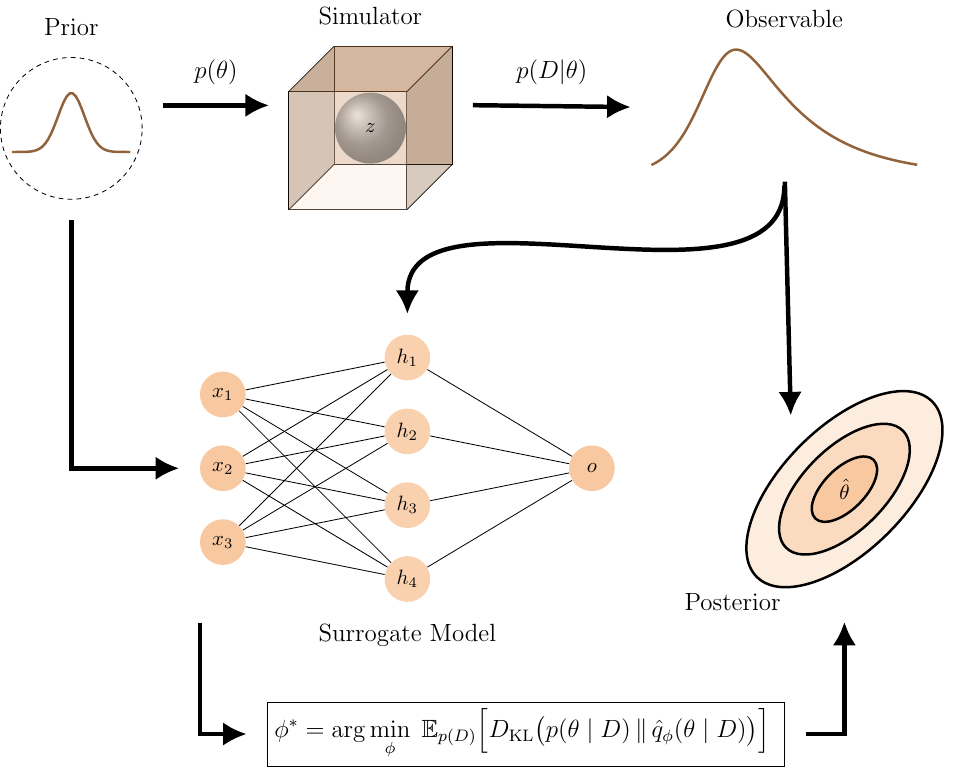}
    \caption{A diagram outlining our amortized inference procedure. }
    \label{fig:amortized}
\end{figure}

Figure \ref{fig:amortized} illustrates this approach. Starting from a prior \(p(\theta)\), a simulator generates mock observations \(D\) based on the latent parameters \(\theta\). These observations are fed into a surrogate model, such as a neural network, which learns to approximate the posterior \(q_\phi(\theta | D)\). Through iterative optimization, the model minimizes the $\mathrm{D_{KL}}$ divergence between the true and approximate posteriors. This framework scales efficiently to large and complex datasets, making it a valuable tool for data-intensive astronomical applications \citep[e.g.][]{karchev2023}.

\subsection*{Surrogate Model}

A common implementation for surrogate models employs \emph{normalizing flows} as neural density estimators to enable efficient, tractable approximations of complex posterior distributions. These models transform a simple base distribution
 \(\pi(z)\), typically a multivariate Gaussian, into the target distribution using an invertible function \(f\). Letting $z = f^{-1}(\theta)$, the transformed density is given by:
\begin{equation}
p(\theta | x) 
= \pi(z) \,\Biggl| \det\!\biggl(\frac{\partial f^{-1}}{\partial \theta}\biggr)\Biggr|.
\end{equation}

\begin{figure}
    \centering
    \includegraphics[width=\columnwidth]{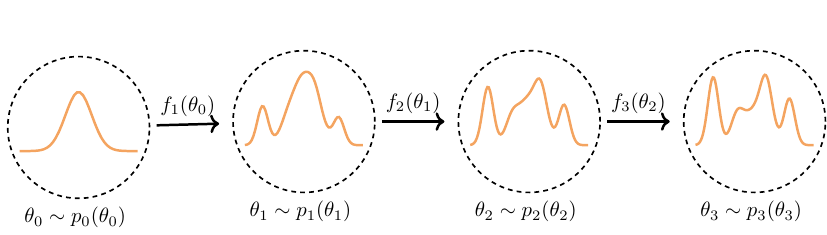}
    \caption{Illustration of a normalizing flow transforming a simple base distribution into a complex target distribution through successive invertible mappings.}
    \label{fig:NF}
\end{figure}

Figure \ref{fig:NF} presents a toy illustration demonstrating how complex distributions can be approximated by applying a sequence of transformations to a Gaussian distribution.
One popular class of normalizing flows is the \emph{Masked Autoregressive Flow}, which models the joint distribution of \(\theta\)
as a product of conditionals:
\begin{equation}
p(\theta) 
= \prod_{i=1}^{d} p(\theta_i | \theta_{1:i-1}),
\end{equation}
where \(d\) is the dimensionality of \(\theta\). By using neural networks to parameterize these conditional distributions, Masked Autoregressive Flow can capture intricate dependencies within the target distribution.

\section*{Active Inference in Cosmology}

Traditional statistical analyses often operate under the assumption that, if more data is needed, it can reasonably be collected. In astronomy, however, this is rarely possible. Our vantage point in the Universe imposes inherent observational limits. As early as 1584, Giordano Bruno noted in \textit{De l’infinito, universo et mondi} that the faintness or distance of celestial bodies could hinder our ability to observe them. This idea was later formalized by Eddington in his 1914 treatise \textit{Stellar Movements and the Structure of the Universe}, where he highlighted that stars visible to the naked eye form a biased sample.  This line of reasoning ultimately led to the concept of \emph{Malmquist bias}: in flux-limited surveys, intrinsically brighter objects are overrepresented, skewing inferences about the true distribution of celestial sources.

Selection effects are particularly evident when comparing spectroscopic and photometric samples, the two main types of observations in  cosmological studies. Spectroscopy involves dispersing an object’s light into its constituent wavelengths to produce a spectrum, which reveals physical properties such as chemical composition, temperature, and velocity. For supernovae, spectroscopy enables precise classification by detecting features such as the absence of hydrogen lines, and the presence of a strong \emph{silicon absorption line}—a dip in the observed light at a specific wavelength, caused by silicon atoms absorbing photons at that energy, which is characteristic of Type Ia events.  In contrast, photometry measures the total brightness of an object through broad filters, capturing less detailed but more easily obtained information. Once a supernova is  classified, its brightness is tracked photometrically over time to estimate its distance. However, spectroscopy is resource-intensive and only feasible for a small subset of objects, whereas photometric surveys can observe far more events at the expense of precision. This imbalance introduces selection effects that can bias cosmological analyses if not properly addressed.

Once the classification is confirmed, the target is continuously monitored using \emph{photometric} observations, which measure the total brightness of the object integrated over broad wavelength ranges, typically using filters such as blue, green, or red bands. The magnitude (or brightness) of the Type Ia supernova at maximum light is the starting point for a cosmological study. These two data types result from different instruments and intrinsically distinct observational strategies.

The spectroscopic subset \(\mathcal{S}_\text{spec} \subseteq \mathcal{S}\) provides higher-resolution information about the spectral properties of objects, whereas the photometric subset \(\mathcal{S}_\text{phot} \subseteq \mathcal{S}\) relies on integrated measurements over a smaller number of wavelength bands. Due to the resource-intensive nature of spectroscopy, \(\mathcal{S}_\text{spec}\) is often smaller and less representative of the underlying population than \(\mathcal{S}_\text{phot}\), introducing systematic biases.

These challenges have prompted a paradigm shift. Rather than passively analyzing fixed datasets, astronomers increasingly embrace \emph{active inference} strategies \citep{Settles:2012, Ishida2019}, where observations are selected strategically to maximize scientific return under limited resources. In this framework, data are no longer static inputs but dynamic choices: their potential informativeness is explicitly considered when deciding which measurements to collect. For example, selecting supernovae for spectroscopic follow-up involves balancing observational cost against the expected scientific gain. The problem resembles a knapsack optimization, where each candidate observation has a different value and must be chosen within the constraints of finite telescope time \citep{kennamer2020}.

In information-theoretic terms, \emph{active inference} aims to minimize ``surprise'' \citep{Parr2022}—
the extent to which new data update prior beliefs. Formally, surprise corresponds to the negative log-evidence, where the evidence is the probability of the observed data under the model, a quantity that is generally intractable. In practice, one minimizes a tractable upper bound: the \emph{variational free energy}.  Within this framework, the goal is to evaluate how much an unobserved measurement would reduce uncertainty about latent variables or parameters of interest—such as redshift, supernova type, or cosmological parameters. A common strategy is to select the next observation \(y\) that minimizes the expected discrepancy between an approximate  posterior \(Q(x)\) and the true, but intractable, posterior.

This strategy can be formalized by minimizing the \emph{variational free energy}:
\begin{equation}
F[Q, y] 
= - \mathbb{E}_{Q(x)} \bigl[\ln P(y, x)\bigr]
- H\bigl[Q(x)\bigr],
\end{equation}
where \(x\) denotes latent variables, such as cosmological parameters, \(y\) the observed data, \(Q(x)\) an approximate posterior over \(x\), and \(H[Q(x)]\) its entropy. The joint distribution \(P(y, x)\) defines the generative model that links observations to latent variables. Observations associated with higher expected surprise, quantified via a greater mismatch between model predictions and data, induce larger posterior updates, and therefore greater information gain.

\begin{figure}
    \centering
    \includegraphics[width=\columnwidth]{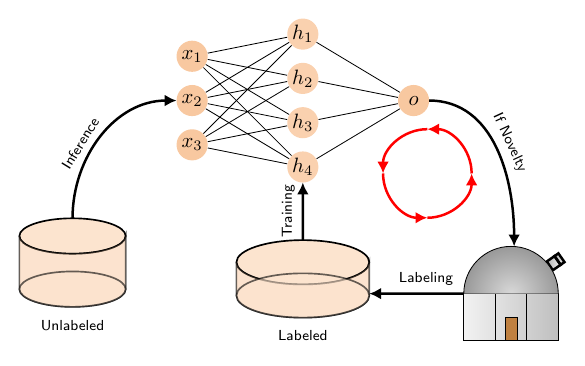}
    \caption{This workflow illustrates an active learning loop in astronomy, where spectroscopic follow-up is prioritized for novel objects that maximize information gain.}
    \label{fig:SNeAL}
\end{figure}

As illustrated in Figure~\ref{fig:SNeAL}, this iterative process combines observational constraints with machine learning to prioritize spectroscopic follow-ups that most effectively reduce uncertainty and correct for selection biases. Over time, such strategies enhance model performance while ensuring that limited observational resources are used efficiently.

\section*{Remarks}

From charting planetary motions by hand to modeling the expansion of the Universe with billions of data points, astronomy has evolved into a discipline where inference is not just a method, but its very foundation.  This journey, from Tycho Brahe’s systematic use of empirical observations to today’s simulation-based and active inference techniques, reflects a deeper shift: from passively recording the heavens to actively interrogating them. This transformation offers immense opportunities, but also demands unprecedented integration of mathematics, statistics, computer science, and physics to advance our understanding of the cosmos.

A testament to this progress is the statistical inference of millions—and even billions—of parameters, once unimaginable. The Gaia satellite, which offers the most comprehensive catalog of stellar positions and motions to date, exemplifies this feat. Its dataset is used to infer tens of billions of astrometric parameters from trillions of time measurements, integrated through a carefully designed mathematical optimization framework. This approach extends the weighted least-squares method, employing a hybrid solution that integrates iterative optimization techniques—blending Jacobi's and Gauss-Seidel block methods—with a modified conjugate gradient method \citep{2012A&A...538A..78L}.
The result is a monumental dataset detailing the positions and motions of billions of stars, which now forms the bedrock of contemporary astronomical research.

The future of astronomical inference stands poised for profound transformation through the integration of advanced statistical models. These models, capable of discerning intricate patterns within vast and complex datasets, promise to unlock new pathways for knowledge discovery while significantly streamlining labor-intensive data processing tasks. As these tools continue to evolve, they will enhance the synergy between human expertise and computational power, allowing astronomers to focus on high-level guidance and interpretation while automation handles much of the underlying analysis. This shift will not only accelerate the pace of discovery, but will also deepen our understanding of the cosmos.

\bibliographystyle{mnras}
\bibliography{Refs}

\end{document}